\documentclass[sigconf]{acmart} 
\usepackage{threeparttable}
\usepackage{multirow}
\newcommand{\eat}[1]{}

\AtBeginDocument{%
  \providecommand\BibTeX{{%
    \normalfont B\kern-0.5em{\scshape i\kern-0.25em b}\kern-0.8em\TeX}}}

\copyrightyear{2024} 
\acmYear{2024} 
\setcopyright{acmlicensed}
\acmConference[WSDM '24]{Proceedings of the 17th ACM International Conference on Web Search and Data Mining}{March 4--8, 2024}{Merida, Mexico}
\acmBooktitle{Proceedings of the 17th ACM International Conference on Web Search and Data Mining (WSDM '24), March 4--8, 2024, Merida, Mexico}
\acmPrice{15.00}
\acmDOI{10.1145/3616855.3635829}
\acmISBN{979-8-4007-0371-3/24/03}

\begin{document}

\title{Deep Evolutional Instant Interest Network for CTR Prediction in Trigger-Induced Recommendation}

\author{Zhibo Xiao}
\affiliation{%
  \institution{Alibaba Group}
  \city{Hangzhou}
  \state{Zhejiang}
  \country{China}
}
\email{xiaozhibo.xzb@alibaba-inc.com}

\author{Luwei Yang}
\authornote{Corresponding author.}
\affiliation{%
  \institution{Alibaba Group}
  \city{Hangzhou}
  \state{Zhejiang}
  \country{China}
}
\email{luwei.ylw@alibaba-inc.com}

\author{Tao Zhang}
\affiliation{%
  \institution{Alibaba Group}
  \streetaddress{}
  \city{Hangzhou}
  \state{Zhejiang}
  \country{China}
}
\email{selous.zt@alibaba-inc.com}

\author{Wen Jiang}
\affiliation{%
  \institution{Alibaba Group}
  \streetaddress{}
  \city{Hangzhou}
  \state{Zhejiang}
  \country{China}
}
\email{wen.jiangw@alibaba-inc.com}

\author{Wei Ning}
\affiliation{%
  \institution{Alibaba Group}
  \streetaddress{}
  \city{Hangzhou}
  \state{Zhejiang}
  \country{China}
}
\email{wei.ningw@alibaba-inc.com}

\author{Yujiu Yang}
\affiliation{%
  \institution{SIGS, Tsinghua University}
  \streetaddress{}
  \city{Shenzhen}
  \state{Guangdong}
  \country{China}
}
\email{yang.yujiu@sz.tsinghua.edu.cn}

\renewcommand{\shortauthors}{Zhibo Xiao et al.}

\begin{abstract}
The recommendation has been playing a key role in many industries, e.g., e-commerce, streaming media, social media, etc.
Recently, a new recommendation scenario, called Trigger-Induced Recommendation (TIR), where users are able to explicitly express their instant interests via trigger items, is emerging as an essential role in many e-commerce platforms, e.g., Alibaba.com and Amazon. Without explicitly modeling the user's instant interest, traditional recommendation methods usually obtain sub-optimal results in TIR. Even though there are a few methods considering the trigger and target items simultaneously to solve this problem, they still haven't taken into account temporal information of user behaviors, the dynamic change of user instant interest when the user scrolls down and the interactions between the trigger and target items. To tackle these problems, we propose a novel method -- Deep Evolutional Instant Interest Network (DEI2N), for click-through rate prediction in TIR scenarios. Specifically, we design a User Instant Interest Modeling Layer to predict the dynamic change of the intensity of instant interest when the user scrolls down. Temporal information is utilized in user behavior modeling. Moreover, an Interaction Layer is introduced to learn better interactions between the trigger and target items. We evaluate our method on several offline and real-world industrial datasets. Experimental results show that our proposed DEI2N outperforms state-of-the-art baselines. In addition, online A/B testing demonstrates the superiority over the existing baseline in real-world production environments.
\end{abstract}

\ccsdesc[500]{Information systems~Personalization}
\ccsdesc[500]{Information systems~Recommender systems}
\ccsdesc[300]{Information systems~Learning to rank}

\keywords{Recommender Systems; Click-Through Rate Prediction; User Instant Interests; Trigger-Induced Recommendation}


\maketitle

\section{Introduction}

\begin{figure}[tb]
  \centering
  \setlength{\abovecaptionskip}{0.06cm}
    \includegraphics[trim = 30mm 5mm 35mm 10mm, clip, width=0.9\columnwidth]{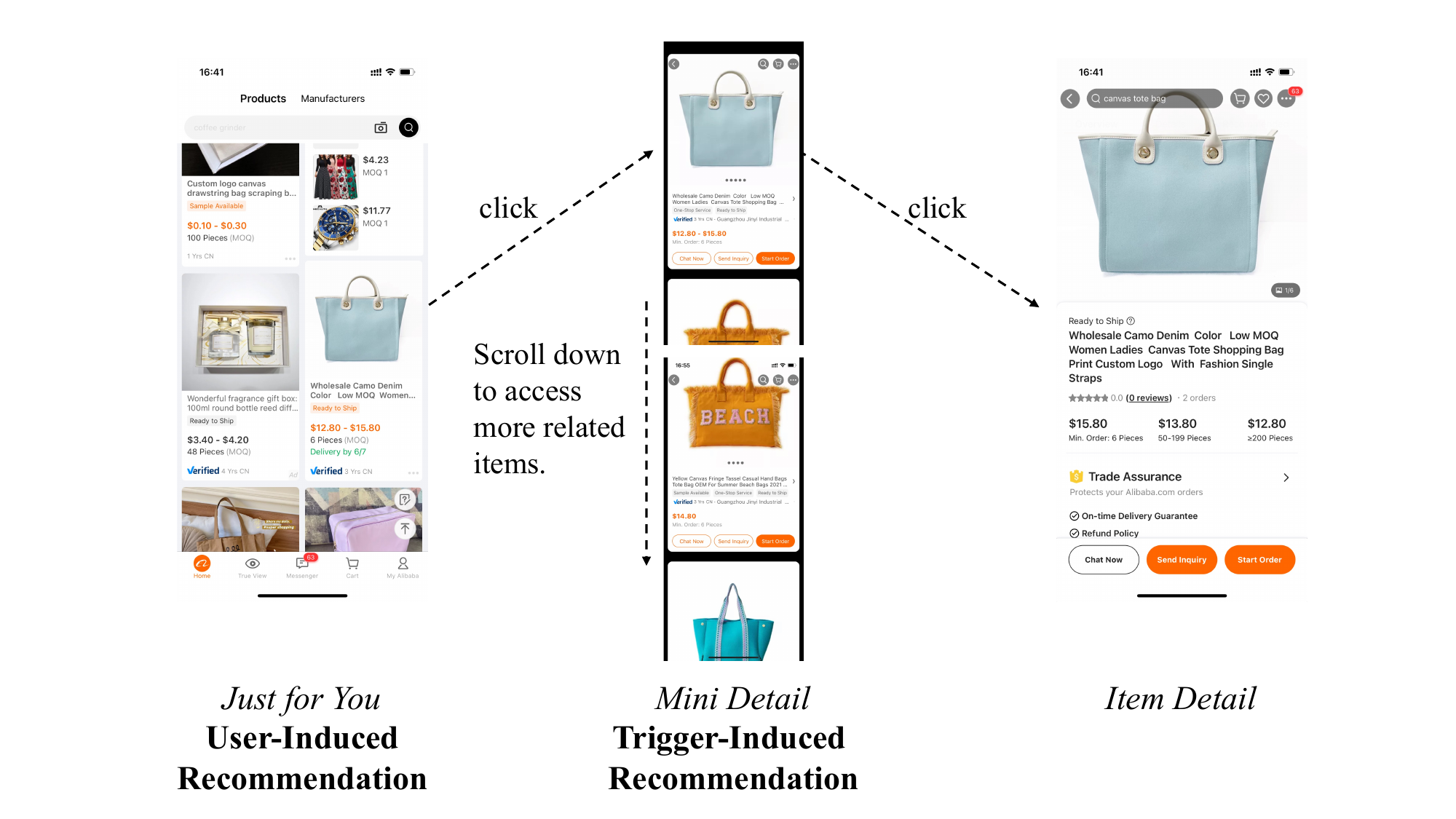}
  \caption{Recommendation scenarios at an e-commerce app. Left: User-Induced Recommendation, middle: Trigger-Induced Recommendation, right: Item Detail.}
  \label{fig:recom_demo}
  \vspace{-1.5em}
\end{figure}

Personalized recommendation systems are extensively employed in the industry. Taking an e-commerce app as an example, we describe two important recommendation scenarios in real industrial platforms applying CTR prediction extensively, User-Induced Recommendation and Trigger-Induced Recommendation, which are shown by Figure~\ref{fig:recom_demo}. The left part shows the \textit{Just for You} module, which is responsible for recommending items according to the user's past interests or behaviors (if permitted by the user). The recommended items in this module are diversified according to the user's historical interests. This scenario is referred to as User-Induced Recommendation (UIR).

Once the user clicks an item, he/she is introduced to a new module, named \textit{Mini Detail}, which is shown in the middle part of Figure~\ref{fig:recom_demo}. Note that, the clicked item in the previous step is presented at the top, which is referred to as the trigger item. The user is able to either click an item to enter the \textit{Item Detail} page (the right part), or scroll down to access more recommended items. These recommended items in \textit{Mini Detail} should be related to the trigger item to some extent. This scenario is often referred to as Trigger-Induced Recommendation (TIR). Besides the \textit{Mini Detail} module, it is very common to see other TIRs, e.g. a \textit{Detail Recommendation} module in \textit{Item Detail} page. Nowadays, TIR is playing an increasingly significant role in many industrial domains, such as e-commerce platforms~\cite{Shen2022DIHN} and messaging APPs~\cite{Xie2021R3S}. In our app, more than $50\%$ of active buyers are contributed by TIR among all recommendation scenarios.

Click-through rate (CTR) prediction plays a crucial role in the recommendation. The main goal is to estimate the likelihood that an item will be clicked by a user. It has a direct and immediate impact on website revenues and user satisfaction, especially in e-commerce. Traditional CTR methods~\cite{Zhou2018DIN,Zhou2019DIEN,Feng2019DSIN,Xiao2020DMIN,Li2020TIEN}, which are more suitable for UIR, have been used extensively in many domains. However, applying it rigidly to TIR could fail to model the instant interest of the user, which results in sub-optimal results.

\begin{figure}[tb]
\setlength{\abovecaptionskip}{0.06cm}
  \centering
    \includegraphics[trim = 2mm 0mm 0mm 12mm, clip, width=0.8\columnwidth]{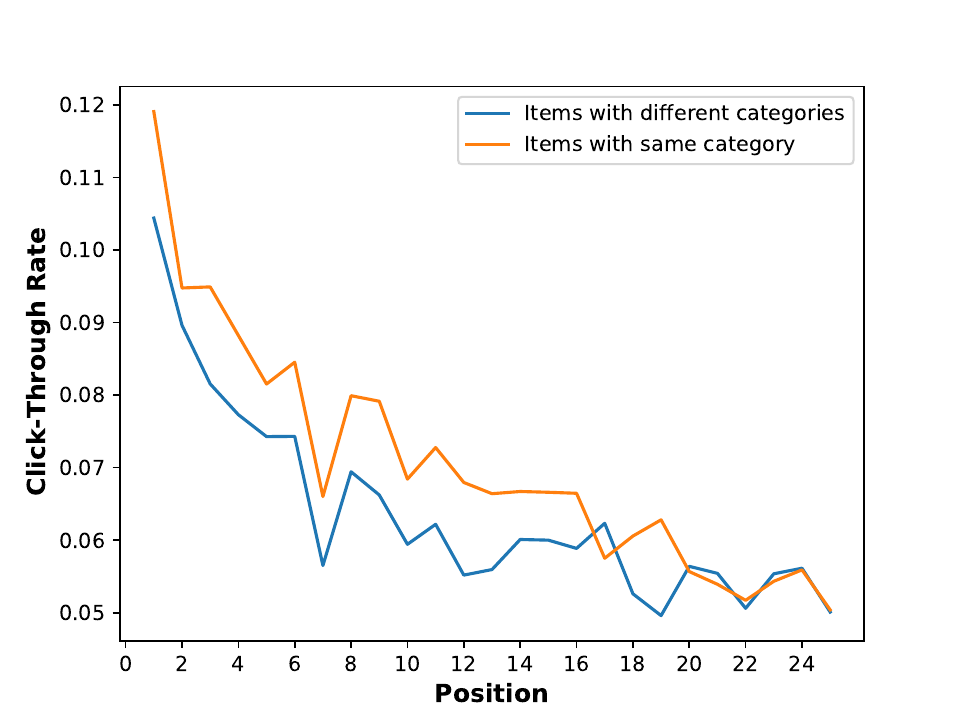}
  \caption{The average click-through rate of items with the same category as the trigger item and items with different categories along the position.}
  \label{fig:CTR_trigger_category}
  \setlength{\belowcaptionskip}{0.9cm}
  \vspace{-1.7em}
\end{figure}

\begin{figure*}[tb]
\setlength{\abovecaptionskip}{0.06cm}
  \centering
  \includegraphics[trim = 4mm 0mm 2mm 17mm,clip, width=1.9\columnwidth]{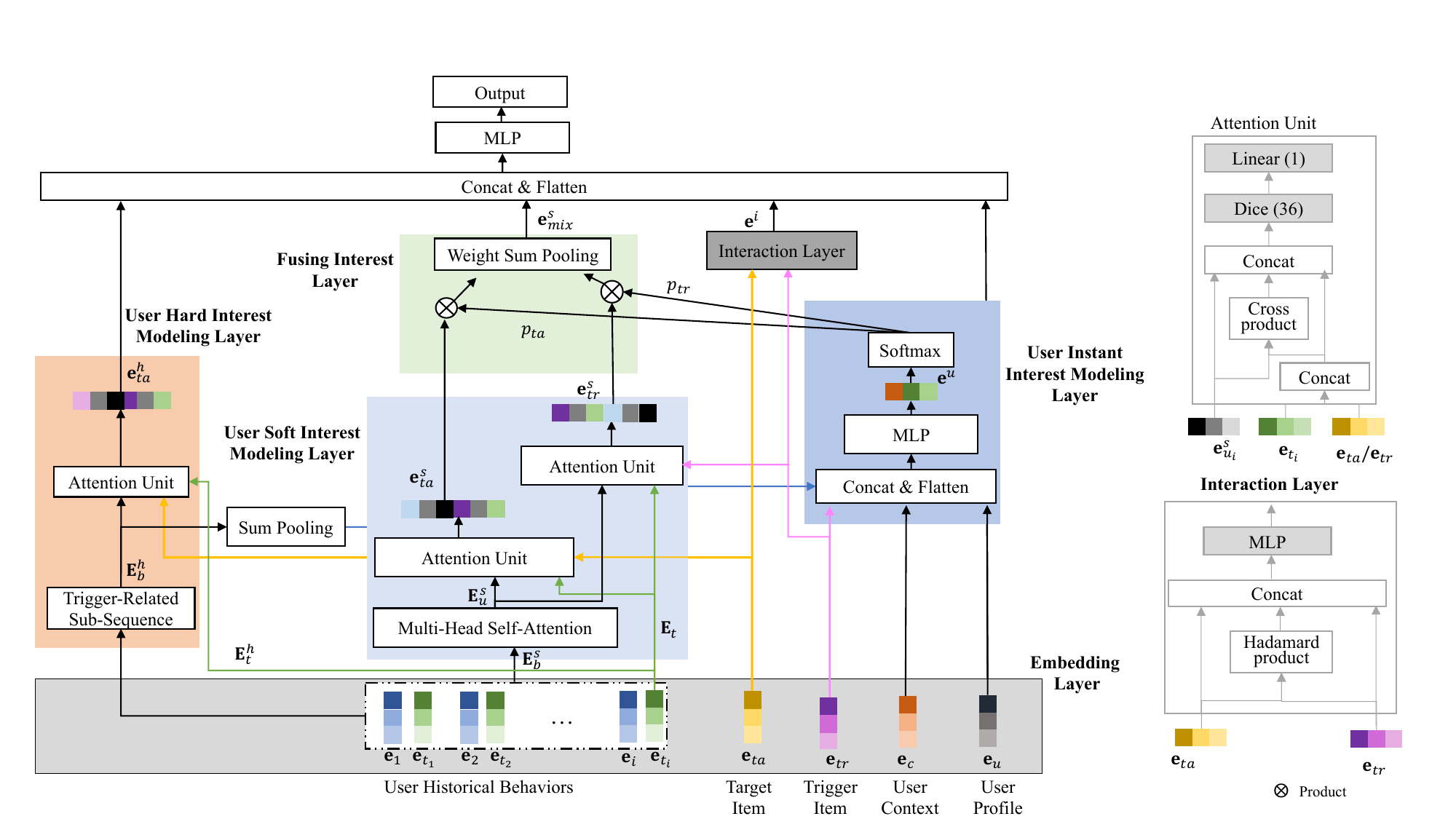}
  \caption{The architecture of the DEI2N model, which consists of Embedding Layer, User Instant Interest Modeling Layer, User Soft Interest Modeling Layer, User Hard Interest Modeling Layer, Fusing Interest Layer and Interaction Layer.}
  \label{fig:DEI2N}
  \vspace{-1.7em}
\end{figure*}

In this paper, we focus on how to accurately estimate the click-through rate of items in Trigger-Induced Recommendation scenarios.
TIR has attracted growing interest in the industry, nonetheless, there is a lack of research on it. R3S~\cite{Xie2021R3S} introduced feature interaction, semantic similarity and information gain to capture users' instant interests. However, it doesn't consider users' historical behaviors, which is one of the most important features in click-through rate prediction modeling. DIHN~\cite{Shen2022DIHN} proposed an interest highlight network to learn the instant interest from the trigger item and the user's historical behaviors. 
DIAN~\cite{Xia2023DIAN} proposed an intent-aware network to learn the user's intention.
However, the temporal information of behaviors, the dynamic change of user instant interest when the user scrolls down and the interactions between the trigger and target items haven't been considered. Specifically, when the user scrolls down, the intensity of instant interest will change dynamically. As shown in Figure~\ref{fig:CTR_trigger_category}, which is based on statistics from a real TIR scene, this phenomenon is confirmed by the \eat{decaying} decreasing gap of CTR between the items with the same category as the trigger item and the items with different categories when the user scrolls down. Therefore, it is highly beneficial to keenly capture the dynamic change of the intensity of instant interest, which is neglected in existing methods~\cite{Shen2022DIHN, Xia2023DIAN}. 


To tackle the aforementioned challenges, we propose a novel method called Deep Evolutional Instant Interest Network (DEI2N~\footnote{The code is released at https://github.com/mengxiaozhibo/DEI2N}) for CTR in TIR scenarios. Specifically, we introduce a User Instant Interest Modeling Layer to predict the dynamic change in the intensity of instant interest when the user scrolls down. This layer is responsible for modeling user instant interest by considering the trigger item and user behaviors simultaneously. Moreover, we integrate temporal information into the attention units to improve the sequence modeling and capture better relevance of the user's interests with respect to the target item and trigger item respectively. Additionally, an Interaction layer is utilized to learn the interaction relationship between the features of the trigger item and target items.

The main contributions of this paper are summarized as follows:
\begin{itemize}
    \item We emphasize an emerging industrial recommendation scenario, Trigger-Induced Recommendation, and highlight the challenges of existing CTR methods applied in TIR.
    \item We propose a novel method DEI2N, which further improves CTR performances in TIR scenarios by considering the dynamic change of user instant interest, temporal information, and the interactions between the trigger and target items. 
    \item We evaluate our method DEI2N on three real-world industrial datasets with state-of-the-art methods. Our method achieves the best performance among competitors. The ablation experiments further verify the effectiveness of the proposed components.
    \item We implement DEI2N in industrial production environments and launch it in five industrial e-commerce TIR scenes. The results of online A/B testing demonstrate the superiority over the existing baseline.
\end{itemize}

\section{Related Work}
As Trigger-Induced Recommendation is an emerging recommendation scenario, we will start with an overview of the CTR prediction task to the most related works in this area. We briefly review three groups of existing methods, 1) Feature Interaction Modeling, 2) User Behavior Modeling, and 3) Trigger-Induced Recommendation. 
 
The first group is \textbf{Feature Interaction Modeling}. Noticing the disability of feature interaction in the linear regression method, a large number of researchers have proposed alternatives to solve this problem. A Factorization Machine model is proposed by~\cite{Rendle2010FM} to model pairwise (second-order) feature interactions. To alleviate the efforts of feature engineering, DeepCrossing~\cite{Shan2016DeepCrossing} utilizes ResNet~\cite{He2016Residual} to automatically learn interactions of features. Wide\&Deep~\cite{Cheng2016WDL} creatively combines the linear model and deep network together for better memorization and generalization. Later, DeepFM~\cite{Guo2017DeepFM} employs a factorization machine instead of a linear model in the wide part.
By noticing the implicit feature interaction of neural networks, DCN~\cite{Wang2017DCN} proposes a more efficient cross-network in addition to a deep network. AutoInt~\cite{Song2019AutoInt} applies the self-attention mechanism to automatically learn feature interactions. Recently, Fi-GNN~\cite{Li2019Fi-GNN}, GraphFM~\cite{Li2021GraphFM}, and DG-ENN~\cite{Guo2021DG-ENN} take advantage of Graph Neural Networks for better feature interactions in feature sparsity scenarios.

The second group is \textbf{User Behavior Modeling}. These methods are more common in the industry. By considering user behaviors, it is able to speculate accurate short-term interests and long-term periodic interests from users. A naive usage of user behavior is just averaging or summing all the behavior embedding vectors to feed into subsequent MLPs. A representative method is DIN~\cite{Zhou2018DIN}. It innovatively uses target attention to obtain the relevant user behaviors respective to the target item. As a result, all relevant user behaviors are activated to calculate the final click-through rate, which in turn obtains better prediction. A subsequent upgraded version DIEN~\cite{Zhou2019DIEN} refines GRUs to model the evolution of user interests from the user behaviors. Considering the multiple interests of user behaviors, DMIN is proposed by \cite{Xiao2020DMIN} to capture the diverse interests of users from his/her behaviors.

As an emerging recommendation scenario, there are few CTR models dedicated to \textbf{Trigger-Induced Recommendation}. Although the above traditional recommendation methods are able to serve in TIR rigidly, the lack of modeling user instant interest motivates researchers to search for better solutions. The most relevant method is DIHN~\cite{Shen2022DIHN}, which is used in an industrial travel e-commerce platform. 
It introduces a user intent network to predict to what extent the user is interested in the trigger item. The output from this network is able to supervise the fusion of interests from the trigger or user behaviors. The second relevant method is DIAN~\cite{Xia2023DIAN}, which uses an intent-ware network to learn the user's intention. The output of this network is used to dynamically balance the results of trigger-free and trigger-based recommendations.
The third most relevant method is R3S~\cite{Xie2021R3S}. It is used in reading recommendation scenarios, where the extended recommendation readings should be relevant to the current clicked reading. The current clicked reading is considered the trigger item in TIR. The recommendations are constructed by taking into account feature interaction, semantic similarity and information gain between the current clicked reading and candidate readings. Nonetheless, the temporal information of behaviors, the dynamic change of user instant interest when the user scrolls down and the interactions between the trigger and target items haven’t been considered.
\section{The Proposed Method}
In this section, we introduce our proposed method, Deep Evolutional Instant Interest Network (DEI2N), for CTR in TIR scenarios. The overall architecture is illustrated by Figure~\ref{fig:DEI2N}. 

We follow the basic CTR paradigm of Embedding \& MLP (Multilayer Perceptron)  model~\cite{Zhou2018DIN}. There are five main components in the middle to better capture user instant interest in TIR. \textit{User Instant Interest Modeling Layer} is responsible for modeling user instant interest by considering the trigger item and user behaviors simultaneously. Additionally, it is able to predict the dynamic change in the intensity of instant interest when the user scrolls down. \textit{User Soft Interest Modeling Layer} and \textit{User Hard Interest Modeling Layer} are applied to extract the user's interests from his/her behaviors according to the trigger and target items. \textit{Fusing Interest Layer} utilizes the results of the User Instant Interest Modeling Layer to fuse the user's interests extracted from the User Soft Interest Modeling Layer. \textit{Interaction Layer} learns the interaction relationship between the features of the trigger item and target items. 
Finally, all of the resulting features and remaining features are concatenated and fed into MLP layers for final CTR prediction. In the remaining section, we will describe these layers in detail.
\subsection{Embedding Layer}
There are five groups of input features: \textit{User Profile}, \textit{User Historical Behaviors}, \textit{Trigger Item}, \textit{Target Item} and \textit{User Context}. 
\textit{User Profile} contains \textit{user ID}, \textit{country ID} and so on. \textit{User Historical Behaviors} is a sequential list of items that the user has clicked or bought. \textit{Trigger Item} and \textit{Target Item} contain \textit{item ID}, \textit{category ID}, \textit{company ID}, etc. \textit{User Context} contains the \textit{page number} that a user is currently browsing. Each feature is normally encoded into a high-dimensional one-hot vector and further is transformed into low dimensional dense features by utilizing embedding layers~\cite{Cheng2016WDL}. 
For example, the \textit{user ID} can be represented by a matrix
$\mathbf{E} \in \mathbb{R}^{K \times {d}_{v}}$, where $K$ is the total number of users and ${d}_{v}$ is the embedding size with $d_{v} \ll K$.
Transformed by embedding layers, \textit{User Profile}, \textit{User Historical Behaviors},  \textit{Trigger Item}, \textit{Target Item} and \textit{User Context} are represented as $\mathbf{e}_{u}$, $\mathbf{E}_{b}$, $\mathbf{e}_{tr}$, $\mathbf{e}_{ta}$ and $\mathbf{e}_{c}$, respectively. Note that, $\mathbf{E}_{b}=\{\mathbf{e}_{1}, \mathbf{e}_{2}, ..., \mathbf{e}_{T}\} \in \mathbb{R}^{T \times {d}_{model} }$, where $T$ represents the length of user historical behaviors and ${d}_{model}$ is the dimension of item embedding $\mathbf{e}_{i}$.

As temporal information is crucial in sequence modeling~\cite{Li2020TIEN}, we introduce the time interval between the historical behavior interaction and the recommendation time. In the formula, the time interval of a historical behavior item is calculated by
${t}_{i}=\lfloor((t-\hat{t}_i)/{T}_{f}) \rfloor$, where $t$ is the recommendation timestamp, $\hat{t}_i$ is the behavior interaction timestamp for item $i$, $\lfloor  \rfloor$ is the floor function,
and 
${{T}_{f}}$ is an adjustable normalization factor. 
We then apply embedding lookup to obtain the time interval embedding $\mathbf{e}_{{t}_{i}}$.
Thus, the time interval representation of the user's historical behaviors is formulated as $\mathbf{E}_{t}=\{\mathbf{e}_{{t}_{1}}, \mathbf{e}_{{t}_{2}}, ..., \mathbf{e}_{{t}_{T}}\} \in \mathbb{R}^{T \times {d}_{time} }$, where ${d}_{time}$ is the dimension of the time interval embedding. 

\subsection{User Instant Interest Modeling Layer}
In TIR scenarios, the clicked trigger item explicitly represents the user's instant interests. Thus, at the beginning, the user is more interested in the items with the same category as the trigger item. However, when the user scrolls down, the intensity of instant interest will change dynamically. This phenomenon is confirmed by the decaying gap of CTR between the items with the same category as the trigger item and the items with different categories when the user scrolls down. Therefore, it is highly beneficial to keenly capture the dynamic change of the intensity of instant interest upon scrolling down, which is neglected in existing methods~\cite{Shen2022DIHN, Xia2023DIAN}.


We propose the User Instant Interest Modeling Layer to predict the dynamic change of the intensity of instant interest upon the user scrolls down. In this layer, we utilize four categories of features, i.e, \textit{User Profile}, \textit{User Context}, \textit{Trigger Item} and the results of sum pooling of the trigger-related sub-sequence as inputs and then feed them into MLPs to generate two probability scores, $p_{tr}$ and $p_{ta}$ with $p_{tr}+p_{ta}=1$. They are formulated as,
\begin{equation}
\label{eq:UI2M_softmax}
{p}_{tr},{p}_{ta} =\textnormal{Softmax}(\textnormal{MLP}(\mathbf{{e}}_{u},\mathbf{{e}}_{c},\mathbf{{e}}_{tr},\textnormal{sum}(\mathbf{{E}}^{h}_{b}))),
\end{equation}
where $\mathbf{{E}}^{h}_{b}$ represents the trigger-related sub-sequence containing the behaviors with the same category as the trigger item. Note that $\mathbf{{E}}^{h}_{b}$ will comprise the most recently interacted item only if this item belongs to the same category as the trigger item. Thus, $p_{tr}$ and $p_{ta}$ represent the extent of how relevant the trigger item and the target item are to user historical behaviors respectively. In other words, it is responsible for determining to what extent the user is interested in the trigger item or target item. Note that $\mathbf{{e}_{c}}$ contains the page number that the user is currently browsing, as we find the page number is a strong signal indicating the evolution of the intensity of user instant interest. 




\subsection{User Soft Interest Modeling Layer}
In traditional CTR prediction methods~\cite{Feng2019DSIN,Zhou2018DIN,Zhou2019DIEN}, user interest modeling is usually implemented by calculating the relevant weights between user historical behaviors and the target item.
However, applying this technique rigidly to TIR would result in non-optimal results. Because the trigger item indicates a strong signal of the user's instant interest. It is inevitable to take both the trigger and target items into account simultaneously.


We propose the User Soft Interest Modeling Layer to extract users' interests with respect to the trigger and target items simultaneously by following~\cite{Shen2022DIHN}. In addition to using Multi-Head Self-Attention (MHSA)~\cite{Vaswani2017Attention} to refine the item representation from user historical behaviors, we introduce residual connection~\cite{He2016Residual}, dropout~\cite{Hinton2012Improving} and layer normalization~\cite{Ba2016LayerNorm} to further improve the item representation. To explicitly introduce temporal information, the input of MHSA is denoted as $\mathbf{E}^{s}_{b}$, which is a concatenation of user historical behavior embeddings $\mathbf{E}_{b}$ and time interval embeddings $\mathbf{E}_{t}$. The MHSA is formulated as:
\begin{equation}\label{eq:Scaled Dot-Product Attention}
\begin{split}
{\mathbf{E}^{s}_{u} = \textnormal{MHSA}(\mathbf{E}^{s}_{b}) = \textnormal{Concat}(\mathbf{head}_{1},\mathbf{head}_{2},...,\mathbf{head}_{{H_R}})\mathbf{W}^{O}},
\end{split}
\end{equation}
\vspace{-1.7em}
\begin{equation}\label{eq:multi_head self_attention_1}
\begin{split}
\mathbf{head}_{h} =&   \textnormal{Attention}(\mathbf{E}^{s}_{b}\mathbf{W}^{Q}_{h},\mathbf{E}^{s}_{b}\mathbf{W}^{K}_{h},\mathbf{E}^{s}_{b}\mathbf{W}^{V}_{h}) \\
=&\textnormal{Softmax}\Big(\frac{\mathbf{E}^{s}_{b}\mathbf{W}^{Q}_{h} \cdot (\mathbf{E}^{s}_{b}\mathbf{W}^{K}_{h})^{\top}}{\sqrt{{d}_{h}}} \Big) \cdot \mathbf{E}^{s}_{b}\mathbf{W}^{V}_{h},
\end{split}
\end{equation}
where $\mathbf{W}^{Q}_{h},\mathbf{W}^{K}_{h},\mathbf{W}^{V}_{h} \in \mathbb{R}^{{d}_{model} \times {d}_{h}}$ are projection matrices of the $h$-th head for query, key and value respectively. The ${H_R}$ is the number of heads and $\mathbf{W}^{O} \in\mathbb{R}^{{d}_{model} \times {d}_{model}}$ is a linear matrix. The $d_{h}$ represents the dimension of each head and $\mathbf{head}_{h}$ represents a latent item representation in subspace.
Next, we apply two attention units to extract the user's interests with respect to the target item and the trigger item separately. Besides and more importantly, the temporal information $\mathbf{E}_{t}$ is utilized in these two attention units to improve the sequence modeling and capture better relevance of the user's interests with respect to the target item and trigger item respectively. Note that, the temporal information is the time interval representation of the user's historical behaviors. The process of the target and trigger attention mechanism can be formulated as:
\begin{equation}
\label{eq:activation_unit_target}
\mathbf{e}^{s}_{ta} =\sum_{j=1}^{T}a(\mathbf{e}^{s}_{{u}_{j}};\mathbf{e}_{{t}_{j}};\mathbf{e}_{ta})\mathbf{e}^{s}_{{u}_{j}}
=\sum_{j=1}^{T}{w}_{{ta}_{j}}\mathbf{e}^{s}_{{u}_{j}},
\end{equation}
\begin{equation}
\label{eq:activation_unit_trigger}
\mathbf{e}^{s}_{tr} = \sum_{j=1}^{T}a(\mathbf{e}^{s}_{{u}_{j}};\mathbf{e}_{{t}_{j}};\mathbf{e}_{tr})\mathbf{e}^{s}_{{u}_{j}}
=\sum_{j=1}^{T}{w}_{{tr}_{j}}\mathbf{e}^{s}_{{u}_{j}},
\end{equation}
where $\mathbf{e}^{s}_{{u}_{j}} \in \mathbb{R}^{{d}_{model}}$ represents the $j$-th item representation after applying MHSA, $\mathbf{e}_{{t}_{j}} \in \mathbb{R}^{{d}_{time}}$ represents the $j$-th item time interval embedding, $a$ is the attention unit which is shown on the top right of Figure~\ref{fig:DEI2N}. 

\subsection{User Hard Interest Modeling Layer}
Motivated by the hard sequential modeling used in SIM~\cite{Pi2020SIM} and DIHN ~\cite{Shen2022DIHN}, we propose the User Hard Interest Modeling Layer. A trigger-related sub-sequence, containing the behaviors with the same category as the trigger item, is aggregated to complement the extraction of users' instant interests. This mechanism helps to filter out irrelevant noise and covers a longer period of user historical behaviors. It is formulated as $\mathbf{E}_{b}^{h}=\{\mathbf{e}^{h}_{{b}_{1}}, \mathbf{e}^{h}_{{b}_{2}}, ..., \mathbf{e}^{h}_{{b}_{{T}_{h}}}\} \in \mathbb{R}^{{T}_{h} \times {d}_{model}}$, where $T_{h}$ is the length of the sub-sequence.
Similarly, the time interval representation of sub-sequence can be formulated as $\mathbf{E}^{h}_{t}=\{\mathbf{e}^{h}_{{t}_{1}}, \mathbf{e}^{h}_{{t}_{2}}, ..., \mathbf{e}^{h}_{{t}_{{T}_{h}}}\} \in \mathbb{R}^{{T}_{h} \times {d}_{time}}$.
Then, we apply the same attention unit used in the previous section to capture the relevance of the user's interests with respect to the target item. Since this sub-sequence is already related to the trigger item, it is not necessary to apply the attention unit with respect to the trigger item. 

Finally, the output of this layer is calculated as,

\begin{equation}
\label{eq:activation_unit_target_hard_sequence}
\mathbf{e}^{h}_{ta} =\sum_{j=1}^{T_h}a(\mathbf{e}^{h}_{{b}_{j}};\mathbf{e}^{h}_{{t}_{j}};\mathbf{e}_{ta})\mathbf{e}^{s}_{{b}_{j}}
=\sum_{j=1}^{T_h}{{w}_{{ta}_{j}}}\mathbf{e}^{s}_{{b}_{j}},
\end{equation}
where $\mathbf{e}^{h}_{{b}_{j}} \in \mathbb{R}^{{d}_{model}}$ represents the $j$-th item representation in trigger-related sub-sequence. The $\mathbf{e}^{h}_{{t}_{j}} \in \mathbb{R}^{{d}_{time}}$ represents the $j$-th item time interval embedding in trigger-related sub-sequence.

\subsection{Fusing Interest Layer}
In order to better model user instant
interest by considering the trigger item, target item and user behaviors simultaneously. We propose the Fusing Interest Layer to utilize the results of the User Instant Interest Modeling Layer to fuse the two user interest representations
extracted from the User Soft Interest Modeling Layer. Mathematically, it is defined as:
\begin{equation}
\label{eq:fusing layer attention}
\mathbf{e}^{s}_{mix} ={p}_{tr}\cdot\mathbf{e}^{s}_{tr}+
{p}_{ta}\cdot\mathbf{e}^{s}_{ta},
\end{equation} 
where ${p}_{tr}, {p}_{ta}$ are the predicted probabilities extracted in the User Instant Interest Modeling Layer and $\mathbf{e}^{s}_{tr}, \mathbf{e}^{s}_{ta}$ are user interest representations extracted in the User Soft Interest Modeling Layer with respect to the trigger and target items respectively.

\subsection{Interaction Layer}
An Interaction Layer, shown on the bottom right of Figure~\ref{fig:DEI2N}, is introduced to learn the explicit interaction relationship between the features of the trigger and target items. It takes the trigger and target items as input and then applies Hadamard product and MLP layers to learn high-order feature interactions,
\begin{equation}
\label{eq:interaction layer}
\mathbf{e}^{i} =\mathrm{MLP}(\mathbf{e}_{ta};\mathbf{e}_{tr};\mathbf{e}_{tr}\times \mathbf{e}_{ta}),
\end{equation}
where $\times$ means the Hadamard product, aka element-wise product.

\subsection{Loss Function}
Finally, all the feature vectors $\mathbf{e}^{s}_{mix}$, $\mathbf{e}^{h}_{ta}$, $\mathbf{e}^{i}$ and $\mathbf{e}_{u}$ are concatenated and then fed into MLP layers for CTR prediction. We adopt the binary cross-entropy loss as the loss function, which is widely used in CTR prediction tasks~\cite{Zhou2018DIN,Zhou2019DIEN,Xiao2020DMIN,Shen2022DIHN},
 \begin{equation}
 \label{eq:cross-entropy loss}
{L} = -\frac{1}{N}\sum_{(\mathbf{x},y)\in S}^{N}\Big({y}\textnormal{log}(f(\mathbf{x}))+ (1-{y})\textnormal{log}(1-
f(\mathbf{x}))\Big),
\end{equation}
where $S$ is the training set of size $N$, $\mathbf{x}$ is the input of network which is a concatenation of $\mathbf{e}^{s}_{mix}$, $\mathbf{e}^{h}_{ta}$, $\mathbf{e}^{i}$ and $\mathbf{e}_{u}$, $y \in \{0,1\}$ is the click label and $f(\mathbf{x})$ is the prediction output of the proposed model.

\section{Evaluation}
\subsection{Datasets}
We use three real-world datasets for evaluation.
One of them is collected from our real-world industrial e-commerce TIR scenario, named Alibaba.com. The other two of them, Alimama and ContentWise, are tailored from existing public e-commerce and media service datasets to fit the TIR problem.
The two of them are sampled from e-commerce platform user logs, and the one among them is obtained from media service. 
The statistics of them are summarized in Table~\ref{tab:dataset}.

\textbf{Alibaba.com.} As there is no public TIR dataset, we create a dataset from an Alibaba.com TIR scenario, \textit{Mini Detail}, which is shown in the middle of Figure~\ref{fig:recom_demo}. The clicked item in the previous step, which is presented at the top in \textit{Mini Detail}, is referred to as the trigger item.
The label is set to positive when a user clicks an item on an exposed list of items in \textit{Mini Detail}, otherwise the label is set to 0.

\textbf{Alimama}~\footnote{https://tianchi.aliyun.com/dataset/dataDetail?dataId=56}. 
To demonstrate the effectiveness of our method, we tailored this dataset to fit the TIR problem by manually creating trigger items.  
We follow the scheme in~\cite{Shen2022DIHN}, the latest clicked item within 4 hours before a sample is deemed as the trigger item. Samples that can not be associated with a trigger will not be selected. 
The label is obtained same as Alibaba.com dataset.

\textbf{ContentWise~}\cite{PerezMaurera2020ContentWise}. 
To evaluate our method on different domains, we introduce a media service dataset which is constructed from an Over-The-Top Media service. The media contents, including movies, movies and clips in series, TV movies or shows, and episodes of TV series, are provided to users by Internet connections. Represented by a recommendation list, the users are able to view the media items, access the media item detail, purchase the media items, or rate the media items.
Since it lacks trigger information, we follow a similar scheme in ~\cite{Shen2022DIHN} to manually create trigger items. Due to the sparsity of the dataset, the latest clicked item within 8 hours before a sample is deemed as the trigger item. Samples that cannot be associated with a trigger will be eliminated. The label is set to positive when a user either views, purchases, rates or accesses the media items. As there is no exposed list items, we follow~\cite{Zhou2019DIEN} to randomly select items as negative samples.


\begin{table}[tb]
\setlength{\abovecaptionskip}{0.05cm}
  \caption{Statistics of the offline datasets.}
  \label{tab:dataset}
  \begin{tabular}{lrrrr}
    \toprule
    Dataset &Users &Items &Categories &Samples\\
    \midrule
    Alibaba.com & 373,852 & 4,715,150 & 6,736 & 5,200,000 \\
    Alimama & 500,000 & 846,812 & 12,978
    & 8,552,702\\
    ContentWise &26,186 & 1,268,988& 117,693 & 2,585,070 \\
  \bottomrule
  \end{tabular}
\setlength{\belowcaptionskip}{0.9cm}
\vspace{-1.5em}
\end{table}

\subsection{Compared Methods}
\label{sec:compared_methods}
To demonstrate the effectiveness of our proposed method, we compare it with several state-of-the-art methods: Wide\&Deep~\cite{Cheng2016WDL}, DIN~\cite{Zhou2018DIN}, DIEN~\cite{Zhou2019DIEN}, DMIN~\cite{Xiao2020DMIN}, DIHN~\cite{Shen2022DIHN} and DIAN~\footnote{The code is not open-sourced, we reproduce it by ourselves.}~\cite{Xia2023DIAN}. 
Besides, we equip some compared methods with the capacity of instant interest modeling for fair and solid comparisons.
\begin{itemize}
    \item \textbf{Wide\&Deep+TR} adds the trigger item as input to capture the user's instant interests.
    \item \textbf{DIN+TRA} applies an attention mechanism to extract the user's instant interests with respect to the trigger item, besides the existing target attention.
    \item \textbf{DIEN+TRA} utilizes the similar attention strategy used in DIN+TRA to better model both the user's instant interest and the user's evolved interest.
    \item \textbf{DMIN+TRA} employs the similar attention strategy used in DIN+TRA to capture the user's instant interest while extracting multiple interests from user historical behaviors. This is the baseline method that we compared in the online A/B testing experiments.
  \vspace{-1.0em}
\end{itemize}

\begin{table*}
\setlength{\abovecaptionskip}{0.15cm}
\centering
\caption{Experimental AUC results on real-world datasets. The bold number in each column indicates the best result, while the underlined number in each column is the second best result.} \label{tab:Experimental Results}
\begin{threeparttable}
  \begin{tabular}{lcccccc}
    \toprule
    \multirow{2}{*}{Model} &\multicolumn{2}{c}{Alibaba.com}  &\multicolumn{2}{c}{Alimama} &\multicolumn{2}{c}{ContentWise}\\
    \cline{2-7}
     &AUC &RelaImpr &AUC &RelaImpr &AUC &RelaImpr \\
    \midrule
    Wide\&Deep 
    &{$0.6096\pm0.0019$} &{$-0.99\%$}
    &{$0.6062\pm0.0008$} &{$-7.97\%$}
    &{$0.9469\pm0.0003$} &{$-7.28\%$}\\
    DIN 
    &{$0.6042\pm0.0016$} &
    {$-5.87\%$} 
    &{$0.6154\pm0.0007$} &
    {$0.00\%$} 
    &{$0.9774\pm0.0002$} &
    {$-0.95\%$}\\
    DIEN
    &{$0.6047\pm0.0025$} &
    {$-5.42\%$} 
    &{$0.6155\pm0.0005$} &
    {$0.09\%$} 
    &{$0.9779\pm0.0013$} &
    {$-0.85\%$} \\
    DMIN
    &{$0.6107\pm0.0011$} &
    {$0.00\%$} 
    &{$0.6154\pm0.0002$} &
    {$0.00\%$} 
    &{$0.9820\pm0.0002$} &
    {$0.00\%$} \\
    \hline 
    Wide\&Deep+TR
    &{$0.7412\pm0.0014$} &
    {$111.89\%$} 
    &{$0.6075\pm0.0018$} &
    {$-6.84\%$} 
    &{$0.9713\pm0.0004$} &
    {$-2.22\%$}\\
    DIN+TRA
    &{$0.7425\pm0.0021$} &
    {$119.06\%$} 
    &{$0.6155\pm0.0015$} &
    {$0.09\%$} 
    &{$0.9803\pm0.0019$} &
    {$-0.35\%$}\\
    DIEN+TRA 
    &{$0.7419\pm0.0019$} &
    {$118.52\%$}
    &{$0.6157\pm0.0004$} &
    {$0.26\%$}
    &{$0.9796\pm0.0015$} &
    {$-0.50\%$}
    \\
    DMIN+TRA
    &{$0.7454\pm0.0007$} &
    {$121.68\%$}
    &{$0.6157\pm0.0003$} &
    {$0.26\%$}
    &\underline{$0.9822\pm0.0003$} &
    \underline{$0.04\%$}
    \\
    \hline
    DIHN
    &{$0.7462\pm0.0006$} &
    {$122.40\%$}
    &{$0.6166\pm0.0008$} &
    {$1.04\%$}
    &{$0.9786\pm0.0012$} &
    {$-0.75\%$}
    \\
    DIAN
    &\underline{$0.7480\pm0.0016$} &
    \underline{$124.03\%$}
    &\underline{$0.6168\pm0.0002$} &
    \underline{$1.21\%$}
    &{$0.9764\pm0.0003$} &
    {$-1.16\%$}
    \\
    \textbf{DEI2N}
    &{$\textbf{0.7671}\pm\textbf{0.0012}$}\tnote{*} & {$\textbf{141.28\%}$}
    &{$\textbf{0.6180}\pm\textbf{0.0005}$}\tnote{*} &
    {$\textbf{2.25\%}$}
    & {$\textbf{0.9840}\pm\textbf{0.0002}$}\tnote{*}
    &{$\textbf{0.41\%}$}
    \\
\bottomrule
\end{tabular}
\begin{tablenotes}
    \item[*] \small{Asterisks represent where DEI2N's improvement over compared methods is significant (one-sided rank-num p-value <0.01).}
\end{tablenotes}
\end{threeparttable}
\vspace{-1.0em}
\end{table*}

\subsection{Parameter Settings}
For parameter setting, ${d}_{model}$, ${d}_{time}$, the dimension of the \textit{user profile} and \textit{user context} are set as 72, 36, 36, and 10, respectively. The learning rate is set as 0.001 and the dropout rate is set as 0.1. The number of heads ${H}_{R}$ used in MHSA is set as 2. The normalization factor ${T}_{f}$ is uniformly set to 60, which means we calculate time interval features in minutes. The maximum length of the user behavior sequence is set as 20, 30, and 30 for Alibaba.com, Alimama and ContentWise, respectively; and the maximum length of user trigger-related behavior sub-sequence as 10, 20, and 10 for Alibaba.com, ContentWise and Alimama, respectively. The hidden layer dimensions of the final MLP align with that of the DIEN model at 200 and 80. Additionally, the MLP in the Interaction Layer employs hidden layers of size 144 and 72, while the User Instant Interest Modeling Layer employs hidden layers with sizes 72 and 36.
The implementations of baselines are acquired from their released repositories. The Grid Search technique is applied to find the optimal hyper-parameters.

\subsection{Performance Comparison}
We use the Area Under ROC (AUC) and RelaImpr as evaluation metrics, which are widely applied in CTR prediction tasks~\cite{Zhou2018DIN,Feng2019DSIN,Xiao2020DMIN,Zhou2019DIEN}.
The experimental results on three real-world datasets are shown in Table~\ref{tab:Experimental Results}.

We find that the traditional methods, namely Wide\&Deep, DIN, DIEN, and DMIN, do not perform well in the TIR scenario, especially in the Alibaba.com dataset. The gaps between the original version and the one equipped with the trigger item are more than 20 percent.
The main reason for this is that these methods do not take the trigger item into account. Once we equip them with the capacity of instant interest modeling, their performances are further improved. These results also show the necessity of elaborate modeling in TIR by considering the trigger item.


 For the sake of fairness, we compare the proposed method DEI2N with DIHN, DIAN and traditional methods equipped by the trigger item. DMIN+TRA is a strong competitor among traditional competitors, which achieves the best results among them. Additionally,
 DIAN, a specialized method for TIR, achieves better results compared with improved versions of traditional methods except for ContentWise. Because it is able to adaptively model both the trigger and target items simultaneously. 
 
 Our proposed method DEI2N obtains the highest AUC value among all state-of-the-art methods.
 The results demonstrate the effectiveness of explicitly considering the dynamic change of user instant interest when the user scrolls down. It allows the model to be aware of the context in order to adaptively fuse user interest representations with respect to the trigger and target items. Besides, modeling of temporal information of user historical behaviors, and the explicit interactions between the trigger and target items contribute to these results as well. 
 We find that the AUC gains of DEI2N over DIAN on Alimama ($0.19\%$) and ContentWise ($0.78\%$) are not obvious as on Alibaba.com ($2.55\%$). One of the possible reasons is that these two datasets are not directly collected from TIR scenarios. The synthesized trigger item may not reflect the real situation in TIR. Furthermore, the lack of context features (e.g., page number) on these two datasets prevents us from modeling the dynamic change of the user's instant interest when the user scrolls down. Consequently, it may limit our model’s performance.

\subsection{Ablation Study}
To understand the effectiveness of the proposed components, we evaluate our proposed method DEI2N in ablation settings. As the Alibaba.com dataset is directly collected from a real-world TIR scenario, we will present ablation results on this dataset. These results  are more realistic and better to show the value of our proposed model.
The ablation experimental results are shown in Table~\ref{tab:Ablation Results}.

To evaluate the effects of the User Instant Interest Modeling layer, we remove this layer as DEI2N-NO-UI2M and compare it with DEI2N. Without explicitly modeling user instant interest, the performance is degraded from an AUC value of 0.7671 to 0.7534. This explicitly shows the benefits of UI2M, which is responsible for predicting the dynamic change of the intensity of instant interest as the user scrolls down. It controls the proportion of the recommended items related to the trigger item. The temporal information on user behaviors is very important. Without the temporal information modeling, DEI2N-NO-TIM degrades the performance from 0.7671 to 0.7652. Temporal information is used in the sequence modeling MHSA and the trigger and targets attention mechanisms in User Soft Interest Modeling Layer and User Hard Interest Modeling Layer. The necessity of explicitly modeling the interactions between the trigger and target items is shown by comparing DEI2N-NO-IL with DEI2N. DEI2N obtains $1.20\%$ relative improvement by introducing explicit interactions between the trigger and target items. The significance of user hard interest and soft interest modeling is represented by the ablation results of DEI2N-NO-UHIM and DEI2N-NO-USIM compared with DEI2N. Without User Hard Interest Modeling Layer and User Soft Interest Modeling Layer, the AUC values are degraded from 0.7671 to 0.7651 and 0.7504 respectively.

\begin{table}[tb]
\setlength{\abovecaptionskip}{0.15cm}
\caption{Ablation experimental results on Alibaba.com dataset.} 
\label{tab:Ablation Results}
\centering
\begin{threeparttable}
  \begin{tabular}{lcc}
    \toprule
    \multirow{2}{*}{Model} &\multicolumn{2}{c}{Alibaba.com}\\
    \cline{2-3}
     &AUC &RelaImpr\\
    \midrule
    DEI2N-NO-UI2M\tnote{a}
    &{$0.7534\pm0.0012$}&{$-5.13\%$}\\
    DEI2N-NO-TIM\tnote{b}
    &{$0.7652\pm0.0013$}&{$-0.71\%$}\\
    DEI2N-NO-IL\tnote{c}
    &{$0.7639\pm0.0008$}&{$-1.20\%$}\\
    DEI2N-NO-UHIM\tnote{d}
    &{$0.7651\pm0.0004$}&{$-0.75\%$}\\
    DEI2N-NO-USIM\tnote{e}
    &{$0.7504\pm0.0010$}&{$-0.7\%$}\\
    \textbf{DEI2N}
    &{$\textbf{0.7671}\pm\textbf{0.0012}$} &{$\textbf{0.00\%}$}
    \\
\bottomrule
\end{tabular}
\begin{tablenotes}
\footnotesize
    \item[a] DEI2N without User Instant Interest Modeling Layer
    \item[b] DEI2N without temporal information modeling
    \item[c] DEI2N without Interaction Layer
    \item[d] DEI2N without User Hard Interest Modeling Layer
    \item[e] DEI2N without User Soft Interest Modeling Layer
\end{tablenotes}
\end{threeparttable}
\setlength{\belowcaptionskip}{-0.9cm}
\vspace{-1.5em}
\end{table}

\subsection{Time Analysis}
In this section, we compare our proposed method with other baselines in training and prediction time. We recorded the training time of these methods on the training set and the test set respectively on the Alibaba.com dataset. The epoch number is set to one. The machine has 41 CPU cores of Intel(R) Xeon(R) Platinum 8163 CPU @ 2.50GHz and 330 GB memory, with 1 NVIDIA Tesla V100 GPU. 

The training and prediction time results are shown in Table~\ref{tab:Execution time}, which indicates the efficiency of DEI2N is comparable to that of these baselines. Because of the extra execution of the trigger item information, all of the traditional baselines are slower than that of their original versions. DIEN+TRA and DMIN+TRA cost 641 and 598 minutes respectively in training, which is the top two of the slowest models. The main reason is that DIEN+TRA uses the GRU module to model the evolution of the user interests and DMIN+TRA has to extract multiple interests from user behaviors. The proposed method DEI2N has almost equivalent time cost as DIHN and DIAN, with 576, 577 and 580 minutes respectively in training. Thus, modeling the temporal information of behaviors, the dynamic change of user instant interest when the user scrolls down, and the interactions between the trigger and target items doesn't introduce much more time cost. The prediction time in the test set has the same tendency.

\begin{table}[tb]
\setlength{\abovecaptionskip}{0.15cm}
  \caption{Execution time (minutes) on Alibaba.com dataset. Training Time is recorded on the training set for one epoch, and prediction time is recorded on the test set for one epoch.}
  \label{tab:Execution time}
  \begin{tabular}{lcc}
    \toprule
    Model &Training Time (m) &Prediction Time (m)\\
    \midrule
    Wide\&Deep &537 & 12 \\
    DIN & 544 & 12 \\
    DIEN & 616 & 15 \\
    DMIN & 581 & 14 \\
    \hline
    Wide\&Deep+TR &540 & 12 \\
    DIN+TRA & 566 & 12 \\
    DIEN+TRA & 641 & 16 \\
    DMIN+TRA & 598 & 14 \\
    \hline
    DIHN & 577 & 13 \\
    DIAN & 580 & 13 \\
    \textbf{DEI2N} & \textbf{576} & \textbf{13} \\
  \bottomrule
  \end{tabular}
  \setlength{\belowcaptionskip}{0.9cm}
   \vspace{-1.5em}
\end{table}

\subsection{Online A/B Testing Results}


Besides the offline performance comparison, we have deployed our proposed method DE2IN in the production environment to do A/B testing. The DEI2N is deployed in Alibaba.com online recommendation systems by leveraging several algorithm platforms in Alibaba Group. 

\begin{figure}[tb]
\setlength{\abovecaptionskip}{0.16cm}
  \centering
  \includegraphics[trim = 30mm 25mm 35mm 10mm, clip, width=1.0\columnwidth]{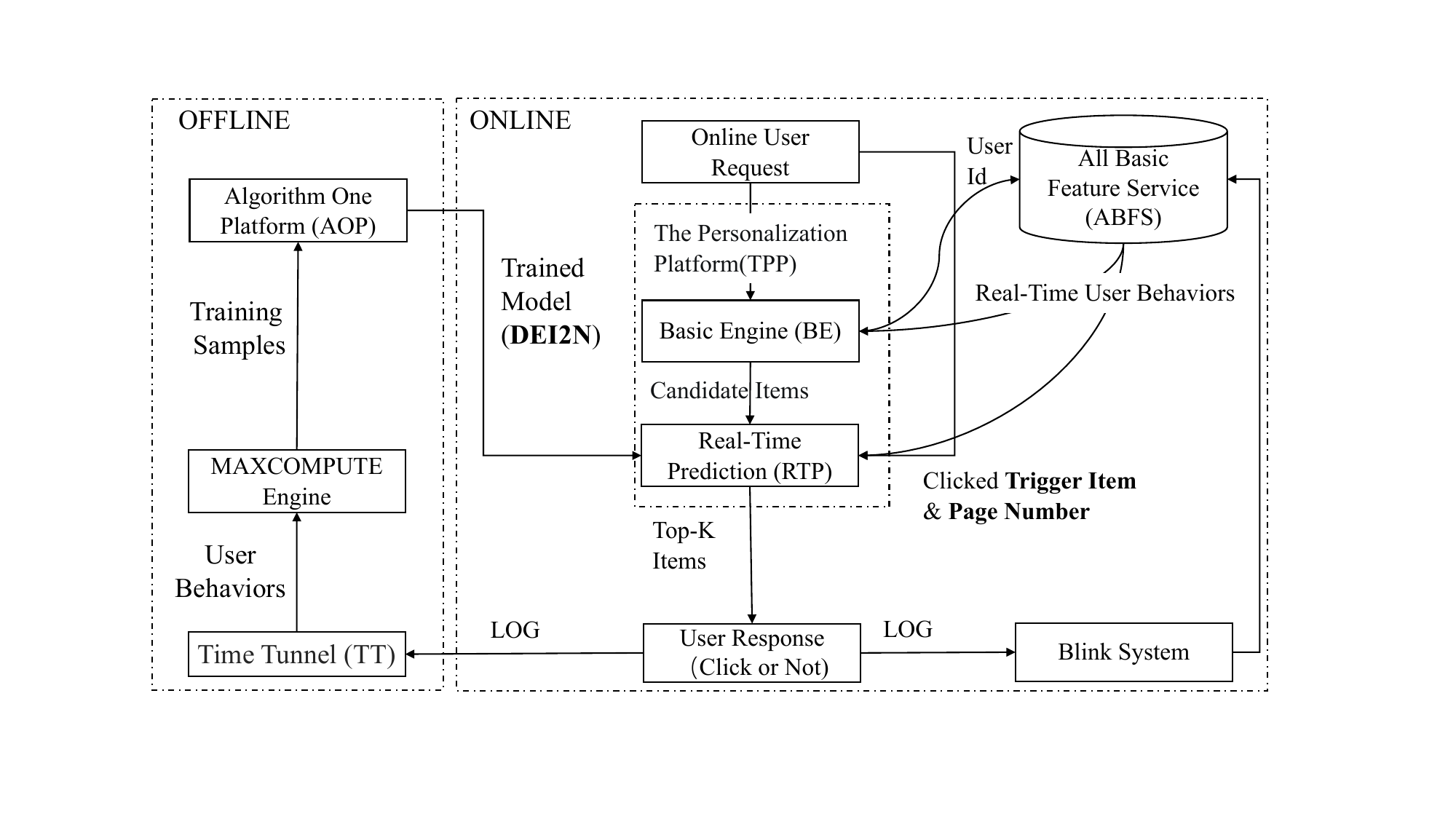}
  \caption{The flowchart of online deployment for DEI2N at Alibaba.com.}
  \label{fig:DEI2N_online_deployment}
  \setlength{\belowcaptionskip}{0.6cm}
   \vspace{-1.5em}
\end{figure}

Figure~\ref{fig:DEI2N_online_deployment} demonstrates the flowchart of online deployment. Basically, there are two main parts in this deployment, online and offline parts. The online part is responsible for generating the final top-k items that will be exposed to end users. Specifically, The Personalization Platform (TPP) accepts real-time request which contains the trigger item and context features such as page number. It then processes the match and rank modules in sequence. The match role is taken by Basic Engine (BE), which will generate thousands of candidate items from tens of millions of candidate item pools. All Basic Feature Service (ABFS) is utilized here to return necessary user features, such as user profile features, real-time user historical behaviors, etc. The rank role is played by Real-Time Prediction (RTP), where our proposed model DEI2N is deployed. It is responsible for calculating CTR scores for the candidate items generated by BE. Then the final top-k items will be exposed to the end user. Note that it is possible to deploy multiple models in RTP, which makes it possible to do A/B testing conveniently. For the offline part, it records the user logs and processes the logs by a big data platform MAXCOMPUTE. Algorithm One Platform (AOP) will accept the processed training samples and train the proposed model DEI2N. Once the training is finished, it will be pushed to RTP for online serving.

We have done A/B testing experiments for several weeks on five different Trigger-Induced Recommendation scenes including \textit{Mini Detail} and \textit{Detail Recommendation}. Considering the facts that DIAN only has a relatively small improvement over DMIN+TRA in offline experiments, usually the efforts to deploy a new model in the production environment are not trivial, and under the business growth pressure, we use the DMIN+TRA as an online baseline model which already has been deployed online. 
DEI2N improves the conversion rate, which is the most important metric in our recommendation scene, by $1.31\%$, $0.56\%$, $1.53\%$, $1.13\%$, and $0.89\%$ for five scenes respectively. These improvements are statistically significant by using an unpaired t-test. It is worth mentioning that the online average response time between DEI2N and DMIN+TRA are almost the same.
Thus DEI2N has been launched in all of the above TIR scenes serving millions of users every day. It demonstrates the effectiveness of DEI2N in real and scalable production environments.

\section{Case Study}
\begin{figure}[tb]
\setlength{\abovecaptionskip}{0.16cm}
  \centering
  \includegraphics[trim = 10mm 0mm 00mm 0mm, clip, width=0.9\columnwidth]{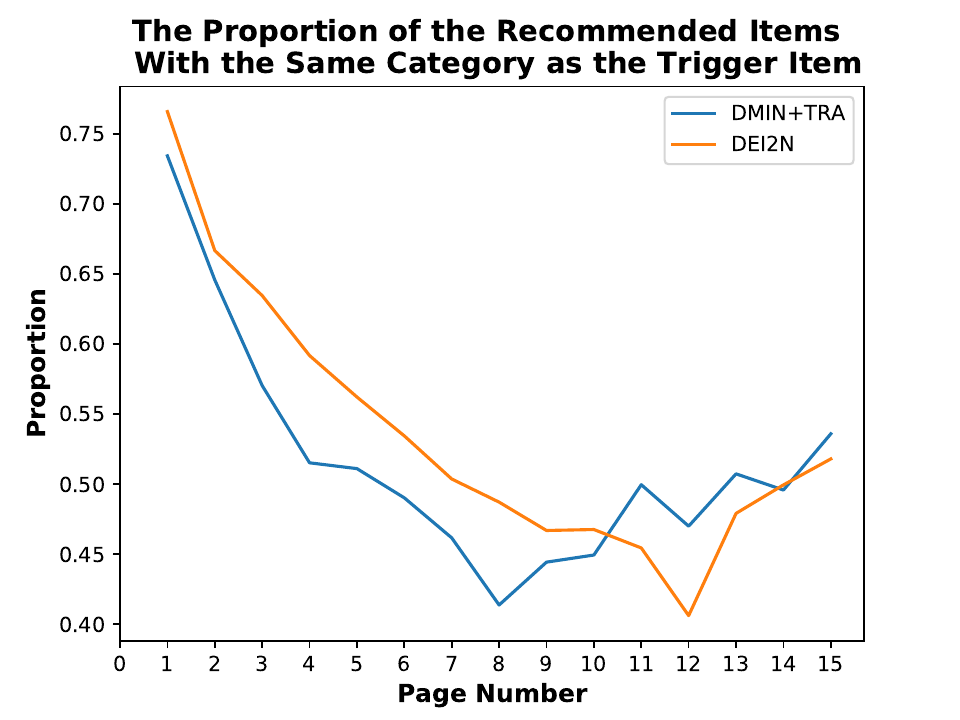}
  \caption{The proportion of the recommended items with the same category as the trigger item along the page number.}
 \setlength{\belowcaptionskip}{0.16cm}
  \label{fig:proportion_same_category}
\end{figure}

Figure~\ref{fig:proportion_same_category} shows the proportion of the recommended items with the same category as the trigger item along the page number. The proportion from DEI2N declined more slowly than the baseline when the page number increased (when the user scrolls down). Thus DEI2N maintains the dynamic change of the user's instant interest more gently, which indicates better modeling of the user's instant interest evolution.

When the user scrolls down (shown as the page number increasing in Figure~\ref{fig:proportion_same_category}), the proportion of the recommended items with the same category as trigger item is dropping down, which means that more diverse items are recommended. 

\begin{figure}[tb]
\vspace{-1.5em}
\setlength{\abovecaptionskip}{0.16cm}
  \centering
  \includegraphics[trim = 0mm 15mm 0mm 00mm, clip, width=\columnwidth]{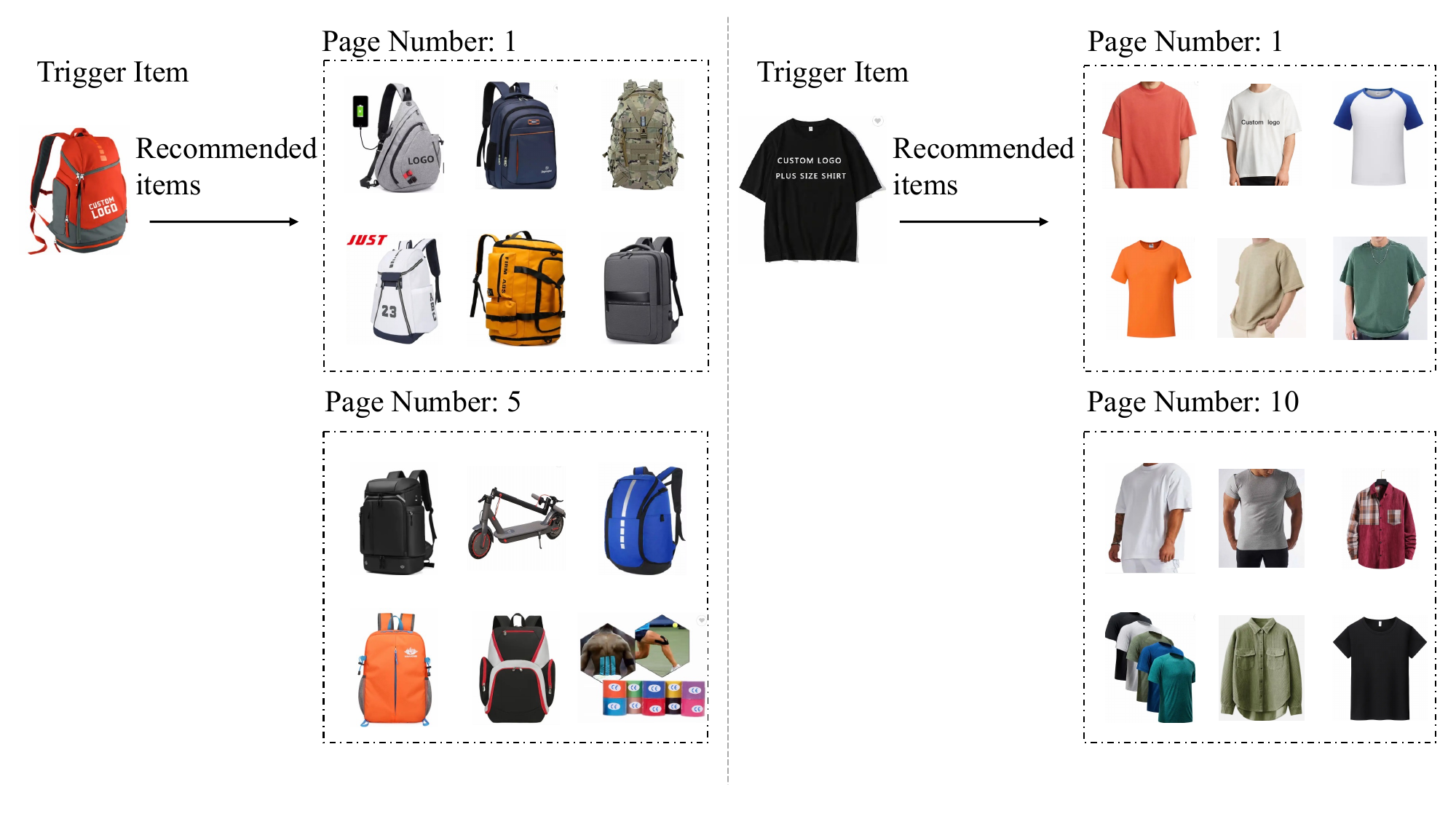}
  \caption{Two cases of DEI2N when the user scrolls down.}
  \label{fig:DEI2N_case_study}
   \setlength{\belowcaptionskip}{0.6cm}
  \vspace{-1.5em}
\end{figure}

We present two detailed user cases of DEI2N as shown by Figure~\ref{fig:DEI2N_case_study}. The left part shows a user coming into our TIR scenario by clicking a sports/gym bag. Thus the clicked bag is the trigger item. The first page shows a large proportion of sports or gym-related bags. When the user scrolls down, for example, to the fifth page, more diverse but relevant to the trigger item to some extent items are recommended. Specifically, an electronic scooter and sport muscle tapes are recommended. It makes sense that these two items are sport-related.
The electronic scooter provides a light solution for going to the gym or sports arena, sport muscle tapes are helpful in muscle pain relief and joint support.

The right part of Figure~\ref{fig:DEI2N_case_study} shows another user case by clicking a black T-shirt. On the first page, almost all of the items are T-shirts. When the user scrolls down to the tenth page, diverse items are presented. We can see that there are two long sleeve shirts among the T-shirts. This diversity is attributed to the elaborated modeling of dynamic change of the intensity of instant interests.

\section{Conclusions}
In this paper, we have proposed a novel method, Deep Evolutional Instant Interest Network (DEI2N), to model user instant interest for click-through rate prediction in TIR scenarios. DEI2N applies a User Instant Interest Modeling Layer to predict the dynamic change of the intensity of instant interest when the user scrolls down in order to extract the user's evolutional instant interests. Temporal information is utilized in modeling layers related to user historical behaviors for better user interest representation. An Interaction Layer is used to explicitly learn better interactions between the trigger and target items. 
Offline experimental results show that our proposed DEI2N achieves the best performance among various state-of-the-art methods in CTR prediction tasks. DEI2N has been deployed in real-world industrial production environments, and the results of online A/B testing demonstrate the superiority over the existing baseline. Improving the conversion rate by several percents, DEI2N has been launched in five industrial TIR scenarios. In the future, we will apply graph learning and contrastive learning to model user’s instant interest by considering the trigger item and user historical behaviors simultaneously, and capture better interactions between the trigger and target items.  


\bibliographystyle{ACM-Reference-Format}
\balance
\bibliography{recom_lite}


\begin{thebibliography}{23}


\ifx \showCODEN    \undefined \def \showCODEN     #1{\unskip}     \fi
\ifx \showDOI      \undefined \def \showDOI       #1{#1}\fi
\ifx \showISBNx    \undefined \def \showISBNx     #1{\unskip}     \fi
\ifx \showISBNxiii \undefined \def \showISBNxiii  #1{\unskip}     \fi
\ifx \showISSN     \undefined \def \showISSN      #1{\unskip}     \fi
\ifx \showLCCN     \undefined \def \showLCCN      #1{\unskip}     \fi
\ifx \shownote     \undefined \def \shownote      #1{#1}          \fi
\ifx \showarticletitle \undefined \def \showarticletitle #1{#1}   \fi
\ifx \showURL      \undefined \def \showURL       {\relax}        \fi
\providecommand\bibfield[2]{#2}
\providecommand\bibinfo[2]{#2}
\providecommand\natexlab[1]{#1}
\providecommand\showeprint[2][]{arXiv:#2}

\bibitem[Ba et~al\mbox{.}(2016)]%
        {Ba2016LayerNorm}
\bibfield{author}{\bibinfo{person}{Jimmy~Lei Ba}, \bibinfo{person}{Jamie~Ryan Kiros}, {and} \bibinfo{person}{Geoffrey~E Hinton}.} \bibinfo{year}{2016}\natexlab{}.
\newblock \showarticletitle{Layer normalization}. In \bibinfo{booktitle}{\emph{arXiv preprint arXiv:1607.06450}}.
\newblock


\bibitem[Cheng et~al\mbox{.}(2016)]%
        {Cheng2016WDL}
\bibfield{author}{\bibinfo{person}{Heng-Tze Cheng}, \bibinfo{person}{Levent Koc}, \bibinfo{person}{Jeremiah Harmsen}, \bibinfo{person}{Tal Shaked}, \bibinfo{person}{Tushar Chandra}, \bibinfo{person}{Hrishi Aradhye}, \bibinfo{person}{Glen Anderson}, \bibinfo{person}{Greg Corrado}, \bibinfo{person}{Wei Chai}, \bibinfo{person}{Mustafa Ispir}, {et~al\mbox{.}}} \bibinfo{year}{2016}\natexlab{}.
\newblock \showarticletitle{Wide \& deep learning for recommender systems}. In \bibinfo{booktitle}{\emph{the 1st workshop on deep learning for recommender systems}}. \bibinfo{pages}{7--10}.
\newblock


\bibitem[Feng et~al\mbox{.}(2019)]%
        {Feng2019DSIN}
\bibfield{author}{\bibinfo{person}{Yufei Feng}, \bibinfo{person}{Fuyu Lv}, \bibinfo{person}{Weichen Shen}, \bibinfo{person}{Menghan Wang}, \bibinfo{person}{Fei Sun}, \bibinfo{person}{Yu Zhu}, {and} \bibinfo{person}{Keping Yang}.} \bibinfo{year}{2019}\natexlab{}.
\newblock \showarticletitle{Deep session interest network for click-through rate prediction}. In \bibinfo{booktitle}{\emph{Proceedings of the 28th International Joint Conference on Artificial Intelligence (IJCAI)}}.
\newblock


\bibitem[Guo et~al\mbox{.}(2017)]%
        {Guo2017DeepFM}
\bibfield{author}{\bibinfo{person}{Huifeng Guo}, \bibinfo{person}{Ruiming Tang}, \bibinfo{person}{Yunming Ye}, \bibinfo{person}{Zhenguo Li}, {and} \bibinfo{person}{Xiuqiang He}.} \bibinfo{year}{2017}\natexlab{}.
\newblock \showarticletitle{DeepFM: a factorization-machine based neural network for CTR prediction}. In \bibinfo{booktitle}{\emph{Proceedings of the 26th International Joint Conference on Artificial Intelligence (IJCAI)}}.
\newblock


\bibitem[Guo et~al\mbox{.}(2021)]%
        {Guo2021DG-ENN}
\bibfield{author}{\bibinfo{person}{Wei Guo}, \bibinfo{person}{Rong Su}, \bibinfo{person}{Renhao Tan}, \bibinfo{person}{Huifeng Guo}, \bibinfo{person}{Yingxue Zhang}, \bibinfo{person}{Zhirong Liu}, \bibinfo{person}{Ruiming Tang}, {and} \bibinfo{person}{Xiuqiang He}.} \bibinfo{year}{2021}\natexlab{}.
\newblock \showarticletitle{Dual Graph enhanced Embedding Neural Network for CTR Prediction}. In \bibinfo{booktitle}{\emph{Proceedings of the 27th ACM SIGKDD Conference on Knowledge Discovery \& Data Mining (KDD)}}.
\newblock


\bibitem[He et~al\mbox{.}(2016)]%
        {He2016Residual}
\bibfield{author}{\bibinfo{person}{Kaiming He}, \bibinfo{person}{Xiangyu Zhang}, \bibinfo{person}{Shaoqing Ren}, {and} \bibinfo{person}{Jian Sun}.} \bibinfo{year}{2016}\natexlab{}.
\newblock \showarticletitle{Deep residual learning for image recognition}. In \bibinfo{booktitle}{\emph{Proceedings of the 2016 IEEE Conference on Computer Vision and Pattern Recognition (CVPR)}}.
\newblock


\bibitem[Hinton et~al\mbox{.}(2012)]%
        {Hinton2012Improving}
\bibfield{author}{\bibinfo{person}{Geoffrey~E Hinton}, \bibinfo{person}{Nitish Srivastava}, \bibinfo{person}{Alex Krizhevsky}, \bibinfo{person}{Ilya Sutskever}, {and} \bibinfo{person}{Ruslan~R Salakhutdinov}.} \bibinfo{year}{2012}\natexlab{}.
\newblock \showarticletitle{Improving neural networks by preventing co-adaptation of feature detectors}.
\newblock \bibinfo{journal}{\emph{arXiv preprint arXiv:1207.0580}} (\bibinfo{year}{2012}).
\newblock


\bibitem[Li et~al\mbox{.}(2020)]%
        {Li2020TIEN}
\bibfield{author}{\bibinfo{person}{Xiang Li}, \bibinfo{person}{Chao Wang}, \bibinfo{person}{Bin Tong}, \bibinfo{person}{Jiwei Tan}, \bibinfo{person}{Xiaoyi Zeng}, {and} \bibinfo{person}{Tao Zhuang}.} \bibinfo{year}{2020}\natexlab{}.
\newblock \showarticletitle{Deep time-aware item evolution network for click-through rate prediction}. In \bibinfo{booktitle}{\emph{Proceedings of the 29th ACM International Conference on Information \& Knowledge Management (CIKM)}}.
\newblock


\bibitem[Li et~al\mbox{.}(2019)]%
        {Li2019Fi-GNN}
\bibfield{author}{\bibinfo{person}{Zekun Li}, \bibinfo{person}{Zeyu Cui}, \bibinfo{person}{Shu Wu}, \bibinfo{person}{Xiaoyu Zhang}, {and} \bibinfo{person}{Liang Wang}.} \bibinfo{year}{2019}\natexlab{}.
\newblock \showarticletitle{Fi-gnn: Modeling feature interactions via graph neural networks for ctr prediction}. In \bibinfo{booktitle}{\emph{Proceedings of the 28th ACM International Conference on Information and Knowledge Management (CIKM)}}.
\newblock


\bibitem[Li et~al\mbox{.}(2021)]%
        {Li2021GraphFM}
\bibfield{author}{\bibinfo{person}{Zekun Li}, \bibinfo{person}{Shu Wu}, \bibinfo{person}{Zeyu Cui}, {and} \bibinfo{person}{Xiaoyu Zhang}.} \bibinfo{year}{2021}\natexlab{}.
\newblock \showarticletitle{GraphFM: Graph Factorization Machines for Feature Interaction Modeling}.
\newblock \bibinfo{journal}{\emph{arXiv preprint arXiv:2105.11866}} (\bibinfo{year}{2021}).
\newblock


\bibitem[P{\'e}rez~Maurera et~al\mbox{.}(2020)]%
        {PerezMaurera2020ContentWise}
\bibfield{author}{\bibinfo{person}{Fernando~B P{\'e}rez~Maurera}, \bibinfo{person}{Maurizio Ferrari~Dacrema}, \bibinfo{person}{Lorenzo Saule}, \bibinfo{person}{Mario Scriminaci}, {and} \bibinfo{person}{Paolo Cremonesi}.} \bibinfo{year}{2020}\natexlab{}.
\newblock \showarticletitle{Contentwise impressions: An industrial dataset with impressions included}. In \bibinfo{booktitle}{\emph{Proceedings of the 29th ACM International Conference on Information \& Knowledge Management (CIKM)}}. \bibinfo{pages}{3093--3100}.
\newblock


\bibitem[Pi et~al\mbox{.}(2020)]%
        {Pi2020SIM}
\bibfield{author}{\bibinfo{person}{Qi Pi}, \bibinfo{person}{Guorui Zhou}, \bibinfo{person}{Yujing Zhang}, \bibinfo{person}{Zhe Wang}, \bibinfo{person}{Lejian Ren}, \bibinfo{person}{Ying Fan}, \bibinfo{person}{Xiaoqiang Zhu}, {and} \bibinfo{person}{Kun Gai}.} \bibinfo{year}{2020}\natexlab{}.
\newblock \showarticletitle{Search-based user interest modeling with lifelong sequential behavior data for click-through rate prediction}. In \bibinfo{booktitle}{\emph{Proceedings of the 29th ACM International Conference on Information \& Knowledge Management (CIKM)}}.
\newblock


\bibitem[Rendle(2010)]%
        {Rendle2010FM}
\bibfield{author}{\bibinfo{person}{Steffen Rendle}.} \bibinfo{year}{2010}\natexlab{}.
\newblock \showarticletitle{Factorization machines}. In \bibinfo{booktitle}{\emph{2010 IEEE International Conference on Data Mining (ICDM)}}. \bibinfo{pages}{995--1000}.
\newblock


\bibitem[Shan et~al\mbox{.}(2016)]%
        {Shan2016DeepCrossing}
\bibfield{author}{\bibinfo{person}{Ying Shan}, \bibinfo{person}{T~Ryan Hoens}, \bibinfo{person}{Jian Jiao}, \bibinfo{person}{Haijing Wang}, \bibinfo{person}{Dong Yu}, {and} \bibinfo{person}{JC Mao}.} \bibinfo{year}{2016}\natexlab{}.
\newblock \showarticletitle{Deep crossing: Web-scale modeling without manually crafted combinatorial features}. In \bibinfo{booktitle}{\emph{Proceedings of the 22nd ACM SIGKDD International Conference on Knowledge Discovery and Data Mining (KDD)}}.
\newblock


\bibitem[Shen et~al\mbox{.}(2022)]%
        {Shen2022DIHN}
\bibfield{author}{\bibinfo{person}{Qijie Shen}, \bibinfo{person}{Hong Wen}, \bibinfo{person}{Wanjie Tao}, \bibinfo{person}{Jing Zhang}, \bibinfo{person}{Fuyu Lv}, \bibinfo{person}{Zulong Chen}, {and} \bibinfo{person}{Zhao Li}.} \bibinfo{year}{2022}\natexlab{}.
\newblock \showarticletitle{Deep Interest Highlight Network for Click-Through Rate Prediction in Trigger-Induced Recommendation}. In \bibinfo{booktitle}{\emph{The World Wide Web Conference (WWW)}}.
\newblock


\bibitem[Song et~al\mbox{.}(2019)]%
        {Song2019AutoInt}
\bibfield{author}{\bibinfo{person}{Weiping Song}, \bibinfo{person}{Chence Shi}, \bibinfo{person}{Zhiping Xiao}, \bibinfo{person}{Zhijian Duan}, \bibinfo{person}{Yewen Xu}, \bibinfo{person}{Ming Zhang}, {and} \bibinfo{person}{Jian Tang}.} \bibinfo{year}{2019}\natexlab{}.
\newblock \showarticletitle{Autoint: Automatic feature interaction learning via self-attentive neural networks}. In \bibinfo{booktitle}{\emph{Proceedings of the 28th ACM International Conference on Information and Knowledge Management (CIKM)}}.
\newblock


\bibitem[Vaswani et~al\mbox{.}(2017)]%
        {Vaswani2017Attention}
\bibfield{author}{\bibinfo{person}{Ashish Vaswani}, \bibinfo{person}{Noam Shazeer}, \bibinfo{person}{Niki Parmar}, \bibinfo{person}{Jakob Uszkoreit}, \bibinfo{person}{Llion Jones}, \bibinfo{person}{Aidan~N Gomez}, \bibinfo{person}{{\L}ukasz Kaiser}, {and} \bibinfo{person}{Illia Polosukhin}.} \bibinfo{year}{2017}\natexlab{}.
\newblock \showarticletitle{Attention is all you need}. In \bibinfo{booktitle}{\emph{Advances in neural information processing systems (NeurIPS)}}. \bibinfo{pages}{5998--6008}.
\newblock


\bibitem[Wang et~al\mbox{.}(2017)]%
        {Wang2017DCN}
\bibfield{author}{\bibinfo{person}{Ruoxi Wang}, \bibinfo{person}{Bin Fu}, \bibinfo{person}{Gang Fu}, {and} \bibinfo{person}{Mingliang Wang}.} \bibinfo{year}{2017}\natexlab{}.
\newblock \showarticletitle{Deep \& cross network for ad click predictions}. In \bibinfo{booktitle}{\emph{Proceedings of the ADKDD'17}}. \bibinfo{pages}{1--7}.
\newblock


\bibitem[Xia et~al\mbox{.}(2023)]%
        {Xia2023DIAN}
\bibfield{author}{\bibinfo{person}{Yaxian Xia}, \bibinfo{person}{Yi Cao}, \bibinfo{person}{Sihao Hu}, \bibinfo{person}{Tong Liu}, {and} \bibinfo{person}{Lingling Lu}.} \bibinfo{year}{2023}\natexlab{}.
\newblock \showarticletitle{Deep Intention-Aware Network for Click-Through Rate Prediction}. In \bibinfo{booktitle}{\emph{The World Wide Web Conference (WWW)}}.
\newblock


\bibitem[Xiao et~al\mbox{.}(2020)]%
        {Xiao2020DMIN}
\bibfield{author}{\bibinfo{person}{Zhibo Xiao}, \bibinfo{person}{Luwei Yang}, \bibinfo{person}{Wen Jiang}, \bibinfo{person}{Yi Wei}, \bibinfo{person}{Yi Hu}, {and} \bibinfo{person}{Hao Wang}.} \bibinfo{year}{2020}\natexlab{}.
\newblock \showarticletitle{Deep multi-interest network for click-through rate prediction}. In \bibinfo{booktitle}{\emph{Proceedings of the 29th ACM International Conference on Information \& Knowledge Management (CIKM)}}.
\newblock


\bibitem[Xie et~al\mbox{.}(2021)]%
        {Xie2021R3S}
\bibfield{author}{\bibinfo{person}{Ruobing Xie}, \bibinfo{person}{Rui Wang}, \bibinfo{person}{Shaoliang Zhang}, \bibinfo{person}{Zhihong Yang}, \bibinfo{person}{Feng Xia}, {and} \bibinfo{person}{Leyu Lin}.} \bibinfo{year}{2021}\natexlab{}.
\newblock \showarticletitle{Real-time Relevant Recommendation Suggestion}. In \bibinfo{booktitle}{\emph{Proceedings of the 14th ACM International Conference on Web Search and Data Mining (WSDM)}}.
\newblock


\bibitem[Zhou et~al\mbox{.}(2019)]%
        {Zhou2019DIEN}
\bibfield{author}{\bibinfo{person}{Guorui Zhou}, \bibinfo{person}{Na Mou}, \bibinfo{person}{Ying Fan}, \bibinfo{person}{Qi Pi}, \bibinfo{person}{Weijie Bian}, \bibinfo{person}{Chang Zhou}, \bibinfo{person}{Xiaoqiang Zhu}, {and} \bibinfo{person}{Kun Gai}.} \bibinfo{year}{2019}\natexlab{}.
\newblock \showarticletitle{Deep interest evolution network for click-through rate prediction}. In \bibinfo{booktitle}{\emph{Proceedings of the AAAI Conference on Artificial Intelligence (AAAI)}}, Vol.~\bibinfo{volume}{33}. \bibinfo{pages}{5941--5948}.
\newblock


\bibitem[Zhou et~al\mbox{.}(2018)]%
        {Zhou2018DIN}
\bibfield{author}{\bibinfo{person}{Guorui Zhou}, \bibinfo{person}{Xiaoqiang Zhu}, \bibinfo{person}{Chenru Song}, \bibinfo{person}{Ying Fan}, \bibinfo{person}{Han Zhu}, \bibinfo{person}{Xiao Ma}, \bibinfo{person}{Yanghui Yan}, \bibinfo{person}{Junqi Jin}, \bibinfo{person}{Han Li}, {and} \bibinfo{person}{Kun Gai}.} \bibinfo{year}{2018}\natexlab{}.
\newblock \showarticletitle{Deep interest network for click-through rate prediction}. In \bibinfo{booktitle}{\emph{Proceedings of the 24th ACM SIGKDD International Conference on Knowledge Discovery \& Data Mining (KDD)}}. \bibinfo{pages}{1059--1068}.
\newblock


\end{thebibliography}


\end{document}